\documentclass[12pt]{article}
\usepackage{amsmath}
\usepackage{amssymb}
\usepackage{amsfonts}
\usepackage{graphicx}
\usepackage[utf8]{inputenc}

\usepackage{slashed}       
\usepackage{mathrsfs}      
\usepackage{bbm}           
\usepackage{bbold}

\usepackage[dvipsnames]{xcolor}
\usepackage{fancyhdr}		
\usepackage{lipsum}         
\usepackage{framed}			
\usepackage{cite}          

\usepackage{booktabs}		
\usepackage{nicefrac}		

\usepackage[font=small]{caption}
\usepackage{float}         
\usepackage{subcaption}

\usepackage[margin=2cm]{geometry}
\graphicspath{{figures/}}
\numberwithin{equation}{section}
\usepackage{tocloft}
\usepackage{multicol}
\usepackage{etoolbox}
\usepackage{relsize}
\patchcmd{\thebibliography}
  {\list}
  {\begin{multicols}{2}\smaller\list}
  {}
  {}
\appto{\endthebibliography}{\end{multicols}}

\let\oldenumerate\enumerate
\renewcommand{\enumerate}{
  \oldenumerate
  \setlength{\itemsep}{1pt}
  \setlength{\parskip}{0pt}
  \setlength{\parsep}{0pt}
}

\let\olditemize\itemize
\renewcommand{\itemize}{
  \olditemize
  \setlength{\itemsep}{1pt}
  \setlength{\parskip}{0pt}
  \setlength{\parsep}{0pt}
}

\newcommand{\acro}[1]{\textsc{\MakeLowercase{#1}}}

\renewcommand{\tilde}{\widetilde}   

\DeclareMathOperator{\tr}{Tr}
\DeclareMathOperator{\Det}{Det}

\newcommand{\SU}[1]{\text{\acro{SU}{\footnotesize#1}}}

\newcommand{\UU}[1]{\text{\acro{U}{\footnotesize#1}}}

\newcommand{\email}[1]{\href{mailto:#1}{#1}}
\newenvironment{institutions}[1][2em]{\begin{list}{}{\setlength\leftmargin{#1}\setlength\rightmargin{#1}}\item[]}{\end{list}}

\usepackage[
	colorlinks=true,
	citecolor= green!50!black,
	linkcolor= NavyBlue!75!black,
	urlcolor=green!50!black,
	hypertexnames=false]{hyperref}

\begin{document}
\begin{center}
	
	{\huge \bf Vector Portal Pseudo-Goldstone Dark Matter\par}
	
	\vskip .7cm
	
	{ \bf
		Ian Chaffey}
	\\ \vspace{-.2em}
	{ \tt
		\footnotesize
		\email{ichaf001@ucr.edu}
	}
	
	\vspace{-.2cm}
	
	\begin{institutions}[2.25cm]
		\footnotesize
		{\it
			Department of Physics \& Astronomy,
			University of  California, Riverside,
			{CA} 92521
		}
	\end{institutions}
	
\end{center}


\begin{abstract}
	\noindent
We present a model of pseudo-Goldstone dark matter that interacts through a light vector mediator based on a spontaneously broken $\SU{(2)}$ dark sector. The dark matter mass is induced by the explicit breaking of the dark $\SU{(2)}$ symmetry. A residual global $\UU{(1)}$ symmetry prevents dark matter decay. The behavior of this model is studied under the assumption that the observed dark matter relic abundance is due to thermal freeze-out. We consider self-interaction targets for small scale structure anomalies and the possibility of interacting with the Standard model through the vector mediator.
\end{abstract}

\small
\setcounter{tocdepth}{2}
\footnotesize
\setcounter{tocdepth}{2}
\tableofcontents
\normalsize
\newpage
\normalsize


\section{Introduction}

The search for the microscopic description of dark matter is a highly active area of research, situated at the crossroads of particle physics and cosmology. While a massive non-luminous particle is required to resolve the large scale structure of the universe, many of dark matter's properties remain unknown~\cite{Ade:2015xua,Aghanim:2018eyx}. A fermionic weakly interacting massive particle (\acro{WIMP}) has long been a leading candidate for dark matter. As experimental constraints tighten~\cite{Tan:2016zwf, Akerib:2015rjg, Akerib:2016vxi,
Agnese:2015nto, Hooper:2012sr, Ackermann:2015zua, Bergstrom:2013jra, Cirelli:2013hv}, dark matter candidates beyond the \acro{WIMP} paradigm are considered. One such alternative is to consider additional states beyond the dark matter itself, referred to as a ``dark" or ``hidden" sector~\cite{%
Pospelov:2008zw,
Pospelov:2008jd,
Pospelov:2007mp,
Essig:2013lka,
Alexander:2016aln,
Battaglieri:2017aum
}. Typically, a dark sector consists of a stable dark matter candidate, as well as one or more states that mediate interactions with the visible sector. Often these additional particles are low mass relative to the dark matter. A common framework is the so-called dark photon model where the mediator is a light spin-1 boson. This dark photon may be observable at current and upcoming experiments~\cite{Essig:2013lka, Alexander:2016aln, Battaglieri:2017aum}. Such dark sectors with light mediators automatically admit long-range, velocity-dependent interactions that may resolve several small-scale structure anomalies~\cite{Tulin:2017ara}.

Spin-0 particles have long been considered as a dark matter candidate. Models of spin-0 dark matter which couple directly to the Standard Model through a Higgs~\cite{SILVEIRA1985136,McDonald:1993ex,Burgess_2001,Cohen_2012,Ma_2006,Honorez_2007} or $Z$-boson portal~\cite{Escudero_2016} have been studied extensively. Dark sector models consisting of spin-0 dark matter and a vector~\cite{B_hm_2004,Boehm:2020wbt} or scalar~\cite{Barger_2009} mediator have been considered. The case of pseudo-Goldstone boson dark matter (\acro{PGBDM}) offers an interesting alternative to these well understood theories. Previously, \acro{PGBDM} has been studied in the context of composite Higgs models~\cite{Frigerio:2012uc}. However in contrast a massive pseudo-Goldstone boson (\acro{PGB}) from a softly broken $\UU{(1)}$ symmetry has a vanishing direct detection cross section at zero momentum transfer when scattering off nuclei through a Higgs portal interaction~\cite{Gross:2017dan,Cline:2019okt}, circumventing experimental constraints from direct detection experiments. The dark matter in such models remains stable due to a $\mathbb{Z}_2$ symmetry. These properties have been shown to hold for the fundamental representation of $\SU{(N)}$~\cite{Karamitros:2019ewv}, and have been studied in the context of $B-L$ extensions of the Standard Model~\cite{Okada:2020zxo}.

In this manuscript, we consider \acro{pGbDM} resulting from the spontaneous breaking of a global symmetry group $\SU{(2)}\times\SU{(2)} \to \SU{(2)}$. This symmetry breaking pattern is analogous to chiral symmetry breaking of the flavor symmetry group in which $\SU{(2)}_L\times\SU{(2)}_R = \SU{(2)}_V\times\SU{(2)}_A$ is broken to the vector subgroup. Here we consider two scalar fields in the fundamental and adjoint representations of a dark sector $\SU{(2)}$ which, when charged under the vector subgroup, explicitly breaks the axial symmetry.
The unbroken $\SU{(2)}$ subgroup is gauged and spontaneously broken in two steps. First, $\SU{(2)} \to \UU{(1)}$ at a scale $f$. And second, $\UU{(1)}\to \varnothing$ at a scale $v \ll f$. 
The symmetry structure permits a residual global $\UU{(1)}$ which stabilizes the dark matter, preventing it's decay into massless states.
The model detailed in this manuscript may also be understood as a phase of the model we present in Ref.~\cite{Chaffey:2019fec}. In that work, we developed the first model of  spin-1 dark matter with a massive spin-1 mediator which originates from a non-abelian gauge sector. Small explicit breaking terms give a mass to the pseudo-Goldstone states which respect the residual $\UU{(1)}$ and remain in the spectrum. We consider the phase of the theory in which these pseudo-Goldstones are the lightest particles in the spectrum which respect the residual $\UU{(1)}$.
The resulting spectrum consists of two heavy gauge bosons with mass on the scale $f$, a dark photon with mass on the scale $v$, the two radial modes responsible for \acro{SSB}, and the two \acro{PGB}'s we consider as dark matter candidates.
Unlike previous models of \acro{PGBDM}, the global symmetry in this model is not softly broken and thus the direct detection cross section resulting from a Higgs portal interaction does not vanish at zero momentum transfer. Instead, we consider the novel case in which the \acro{PGBDM} interacts with the Standard Model via the dark photon, which couples to the Standard Model hypercharge gauge bosons through kinetic mixing~\cite{Holdom:1985ag, GALISON1984279}.
This scenario is analogous to charged pion dark matter.
While \acro{PGBDM} which couples to the Standard Model though the Higgs portal has been well studied, the case where \acro{PGBDM} couples to the Standard Model via a spin-1 mediator charged under an approximate $\UU{(1)}$ symmetry has been relatively unexplored. Previous work has considered models of \acro{PGBDM} in which the dark sector symmetry group is a mirror of the \acro{QCD}~\cite{Tsai:2020vpi} or Standard Model~\cite{Hall:2019rld,Hall:2021zsk} symmetry groups. In the former scenario, resonant scattering between dark matter particles may resolve small-scale structure anomalies as well. In this work we consider a symmetry structure unique from the Standard Model gauge group.

\section{Symmetry Structure}

The scalar sector of our theory consists of an $\SU{(2)}$ doublet $H^i$ and a triplet $\Phi=\phi_a T^a$ where $T^a=\sigma^a/2$ are the generators of $\SU{(2)}$ and $\sigma^a$ are the Pauli matrices. In the limit of no interactions, the full symmetry group of the dark sector is 
\begin{align}
\SU{(2)}_{\Phi} \times \SU{(2)}_H \times \UU{(1)}_H=\SU{(2)}_V \times \SU{(2)}_A \times \UU{(1)}_H \label{eq:fullsym}
\end{align}
with the field transformations for the dark sector given by
\begin{align}
&\SU{(2)}_{\Phi} :\Phi \to U_{\Phi} \Phi U_{\Phi}^{\dagger}
&\SU{(2)}_H : H \to U_H H&
&\UU{(1)}_H : H \to e^{i \theta_H} H
\label{eq:transformations}
\end{align}
where $U_{\Phi,H}=\exp{\left(i \alpha_{\Phi,H}^a T^a\right)}$ is a $2 \times 2$ special unitary matrix. By analogy to chiral symmetry breaking, we have expressed the full symmetry group in terms of its so called vector and axial subgroups. The vector subgroup corresponds to the transformations where $U_{\Phi}=U_H$ while the axial subgroup corresponds to the transformations where $U_{\Phi}=U_H^{\dagger}$. We gauge the vector subgroup $\SU{(2)}_V$, explicitly breaking $\SU{(2)}_A$. The global $\UU{(1)}_H$ symmetry corresponds to an accidental ``Higgs number" symmetry and behaves similar to Standard Model hypercharge.

\subsection{Lagrangian and Symmetry Breaking Potential}
The most general, renormalizable Lagrangian which respects $\SU{(2)}_V\times\UU{(1)}_{H}$ is:
\begin{align}
\mathcal{L}=-\frac{1}{4}F^a_{\mu \nu} F^{a \mu \nu}+|DH|^2+\tr |\mathcal{D}\Phi|^2-V(H,\Phi) \label{eq:L}
\end{align}
where $D$ and $\mathcal{D}$ are the covariant derivatives for $\SU{(2)}$ in the fundamental and adjoint representations respectively. The potential $V(H,\Phi)$ is responsible for spontaneously breaking $\SU{(2)}_V \times \UU{(1)}_H \to \UU{(1)}_{H'}$ as well as explicitly breaking $\SU{(2)}_A$. The most general renormalizable potential invariant under $\SU{(2)}_V\times \UU{(1)}_H$ can be written in the form
\begin{align}
V & =
\phantom{+}
\frac{\lambda}{4!}
\left(
  2 \tr \Phi^2 - f_0^2
\right)^2
+
\frac{\lambda'}{4!}
\left(
  2|H|^2 - v_0^2
\right)^2
+ \mu H^{\dagger}\Phi H
+ \lambda'' |H|^2 \tr \Phi^2 \,.
\label{eq:V}
\end{align}
The first term spontaneously breaks $\SU{(2)}_{\Phi} \to \UU{(1)}_{\Phi}$, the second term spontaneously breaks $\SU{(2)}_H \to \varnothing$, and the trilinear term explicitly breaks $\SU{(2)}_A$. The quartic term proportional to $\lambda''$ introduces mixing between the doublet and the triplet. Additional quartic terms can all be reduced to the term proportional to $\lambda''$. We neglect terms involving the pseudo-conjugate field $\tilde H^i \equiv \epsilon^{ij}H^\dag_{\phantom\dag j}$ as they violate the $\UU{(1)}_H$ symmetry which we assume is respected by the Lagrangian. In the following sections we study the symmetry breaking pattern induced by the vacuum expectation values of the scalar fields $H$ and $\Phi$.

\subsection{Spontaneous Symmetry Breaking}

We assume the vacuum expectation values (vevs)
\begin{align}
\langle \Phi \rangle =f T^3&
&\langle H \rangle=\frac{1}{\sqrt{2}}
\begin{pmatrix}
0\\
v
\end{pmatrix}  \label{eq:vevs}
\end{align}
which break $\SU{(2)}_{\Phi}\to \UU{(1)}_{\Phi}$ and $\SU{(2)}_H \to \varnothing$ respectively. All three generators of $\SU{(2)}_H$ are broken by $\langle H \rangle$ while the $\SU{(2)}_{\Phi}$ generator parallel to $\langle \Phi \rangle$ remains unbroken. The resulting spectrum of states consists of two massive radial modes and five massless Goldstone bosons: two corresponding to the broken generators of $\SU{(2)}_{\Phi}$ and three to the broken generators of $\SU{(2)}_H$.

While the $\UU{(1)}_H$ symmetry is spontaneously broken by \eqref{eq:vevs}, the linear combination
\begin{align}
T_{H'}=T^3+\frac{1}{2}T_H \, , \label{eq:THp}
\end{align}
where $T_H$ is the generator of $\UU{(1)}_H$, remains unbroken. In this representation $T_{H}=\mathbb{1}_{2 \times 2}$. We denote the symmetry generated by $T_{H'}$ as $\UU{(1)}_{H'}$. While only the vector subgroup, $\UU{(1)}_V$, spontaneously broken by $\langle H \rangle$ is gauged, the associated dark photon's interactions respect $\UU{(1)}_{H'}$ as well.  We therefore identify the global \textit{charge} of a dark sector state with respect to this symmetry. This residual symmetry ensures the stability of the lightest charged state.

\subsection{Explicit Symmetry Breaking}

By explicitly breaking the global symmetry group, we can remove all of the massless degrees of freedom from the theory. Our dark sector is constructed to contain two sources of explicit global symmetry breaking:
\begin{enumerate}
	\item Gauge bosons from the explicit gauging of the subgroup $\SU{(2)}_V$.
	
	\item Trilinear mixing between the fundamental scalar $H$ and adjoint scalar $\Phi$.
\end{enumerate}
Gauging $\SU{(2)}_V$ explicitly breaks $\SU{(2)}_A$ since the covariant derivatives are constructed to only respect the symmetry of the gauged subgroup. As a result three massless Goldstone modes are eaten by vector bosons, becoming their longitudinal components. The remaining two degrees of freedom become pseudo-Goldstone bosons (\acro{pGB}) associated with $\SU{(2)}_A$. In the absence of explicit symmetry breaking terms in the potential the \acro{pGB}s are massless at tree level. By introducing a trilinear mixing between $H$ and $\Phi$ the \acro{pGB}s pick up a finite mass proportional to the root of the order of parameter responsible for the explicit breaking $\mu$. We can see such a term should be allowed by considering a product of irreducible representations of $\SU{(2)}$:
\begin{align}
\mathbf{2}\otimes \bar{\mathbf{2}}\otimes \mathbf{3}=\mathbf{5}\oplus\mathbf{3}\oplus\mathbf{3}\oplus\mathbf{1}
\, ,
\end{align}
noting that the singlet corresponds to the operator $H^{\dagger}\Phi H$. From \eqref{eq:transformations} we can see this operator is clearly invariant under transformations where $U_H =U_\Phi$, corresponding to an $\SU{(2)}_V$ singlet.  Since $\SU{(2)}_A$ is not a proper subgroup of $\SU{(2)}_{\Phi}\times \SU{(2)}_H$, operators of this form which break $\SU{(2)}_{\Phi}\times \SU{(2)}_H$ may only be singlets under $\SU{(2)}_V$ or break the symmetry completely. By removing the massless degrees of freedom we have ensured the lightest stable particle in our dark sector is indeed massive, and thus a viable dark matter candidate.

\subsection{Particle Spectrum}

\begin{figure}
  \begin{center}
    \includegraphics[width=\textwidth]{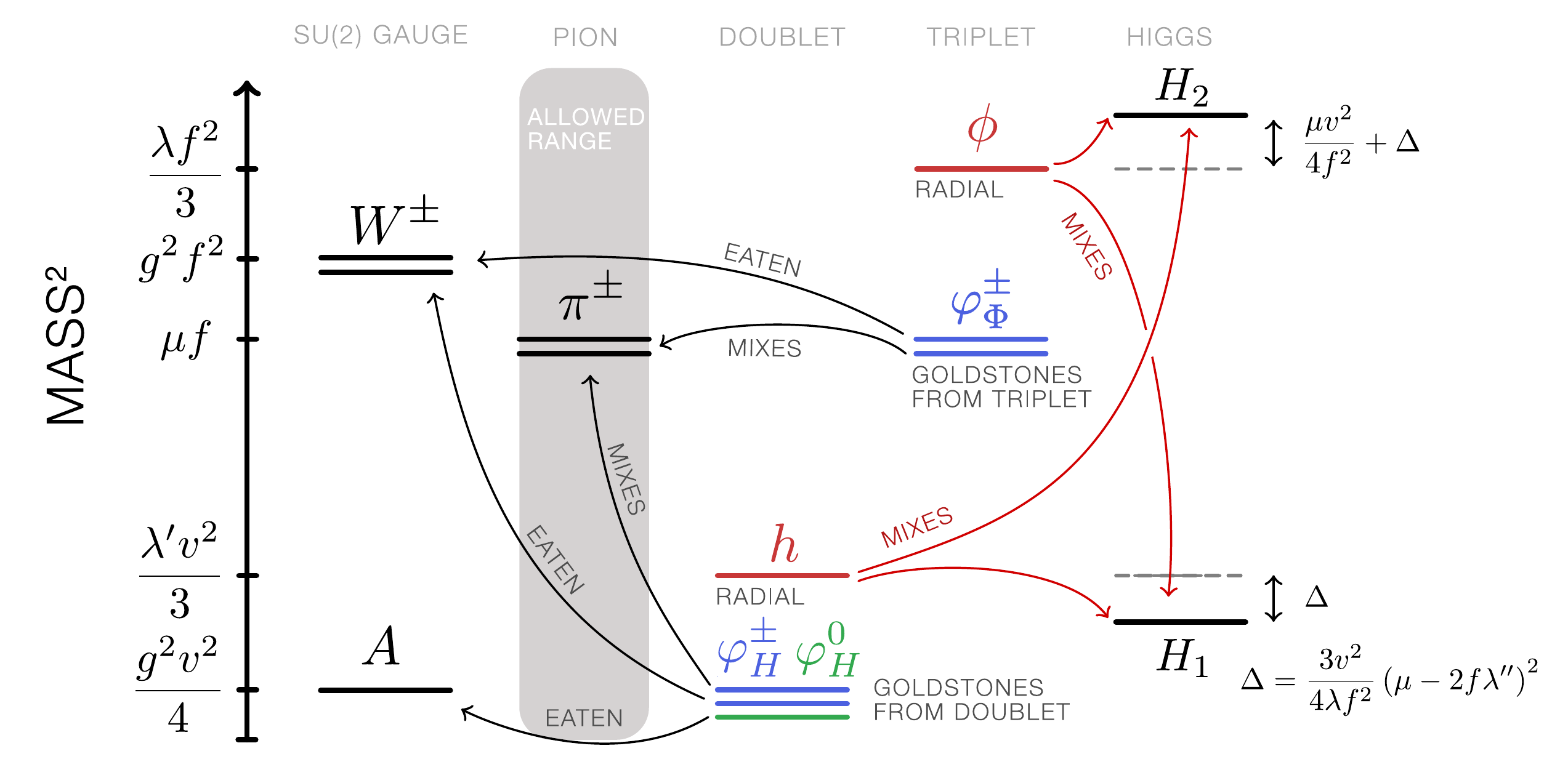}
  \end{center}
  \caption{Model spectrum. Mass eigenstates are black lines, charged (neutral) Goldstones are blue (green) lines, radial Higgs modes are red lines. Mixing into mass eigenstates indicated by thin lines.}
  \label{fig:spectrum}
\end{figure}

Our theory yields a rich spectrum of states. After symmetry breaking, the remaining degrees of freedom consist of three massive gauge bosons, the two scalar radial modes, and two massive \acro{PGB}. We take the limit where
\begin{align}
\langle \tr \Phi^2 \rangle = \frac{f^2}{2}
\qquad\gg\qquad
\langle |H|^2\rangle =\frac{v^2}{2}
\end{align}
setting the mass scale of the gauge boson associated with the broken $\UU{(1)}_V$ subgroup much lower than the gauge bosons corresponding to the other two broken generators. Requiring the mediator to be light and remaining gauge bosons to be heavier than the \acro{PGB}s amounts to the tuning of a dimensionful renormalizable parameter. The resulting particle content can be summarized as follows:
\begin{enumerate}
\item \textbf{Dark Matter}: Pseudo-Goldstone bosons $\pi^\pm$ with mass $\sim \sqrt{\mu f}$.

\item \textbf{Mediators}: Dark photon $A$ and light radial mode $H_1$ with masses $\sim gv$, and $\lambda' v$ respectively.

\item \textbf{Heavy modes}: $W^\pm$ gauge bosons and radial mode $H_2$ with masses $\sim g f$ and $\lambda f$ respectively.
\end{enumerate}
Due to $\UU{(1)}_{H'}$ being unbroken, the lightest charged state remains stable. Since both $W^\pm$ and $\pi^\pm$ are charged under $\UU{(1)}_{H'}$, we assume $m_\pi < m_W+\min(m_1,m_A)\simeq m_W$. The opposite case in which $m_W < m_\pi+\min(m_1,m_A)$ has been studied in Ref.~\cite{Chaffey:2019fec}. We label our fields, wherever possible, in analogy to those of the Standard Model in order to highlight their similar roles in the symmetry structure of our theory.
We sketch the spectrum in Fig.~\ref{fig:spectrum}.

\section{Particles and Mass Spectrum}
Following the CCWZ construction~\cite{PhysRev.177.2247} we parameterize the scalar fields as rotations of the radial modes by the broken generators
\begin{align}
&\Phi=e^{i \frac{\varphi_\Phi \cdot T}{f}}\left(\phi+f\right)T^3 e^{-i\frac{\varphi_\Phi \cdot T}{f}}
&
&H=e^{i \frac{\varphi_H \cdot T}{v/2}}\begin{pmatrix}0 \\ \left(h+v\right)/\sqrt{2}\end{pmatrix}
\end{align}
where we define the broken generators
\begin{align}
&  \varphi_\Phi \cdot T
   =
 \frac{1}{ \sqrt{2}}\varphi_\Phi^+ T^+
  +
 \frac{1}{\sqrt{2}}\varphi_\Phi^- T^-
&
& \varphi_H \cdot T
  =
  \frac{1}{\sqrt{2}}\varphi_H^+T^+
  +
  \frac{1}{\sqrt{2}}\varphi_H^- T^- + \varphi_H^0 T^3
\end{align}
with $T^\pm =T^1 \pm i T^2$. The radial modes $\phi$ and $h$ have been expanded about their respective vevs $f$ and $v$. While the scalar fields do mix with the Standard Model Higgs in principle, we consider this coupling to be negligible throughout this work.

\subsection{Gauge Boson Masses}
The covariant derivatives for the scalar fields are
\begin{align}
&D_{\mu} H=\partial_{\mu} H-i g T^a A^a_{\mu} H
&
&\mathcal{D}_{\mu}=\partial_{\mu} \Phi -i g A_{\mu}^a \left[T^a, \Phi\right]
\end{align}
where $g$ and $A^a_{\mu}$ are the $\SU{(2)}_V$ coupling and gauge field respectively. Evaluating the kinetic terms in the Lagrangian at the vevs gives
\begin{align}
\mathcal{L}\supset g^2\left (f^2+\frac{v^2}{4}\right)W^+_{\mu}W^-_{\mu}+\frac{g^2}{8} v^2 A_{\mu}A^{\mu} \label{eq:Lmass}
\end{align}
where we have defined the mass eigenstates
\begin{align}
& W^{\pm}=\frac{1}{\sqrt{2}}\left(A^1\mp i A^2\right)
&
&A=A^3  \label{eq:masseigenstates}
\end{align}
such that they are labeled according to their $\UU{(1)}_{H'}$ charges. The gauge boson masses follow directly from \eqref{eq:Lmass} to be
\begin{align}
&m_W^2=g^2\left(f^2+\frac{v^2}{4}\right)
&
&m_A^2=\frac{g^2 v^2}{4} \,. \label{eq:gaugebosonmass}
\end{align}

Even before we diagonalize the potential, the massive pseudo-Goldstone eigenstates can be identified from the kinetic terms of the Lagrangian:

\begin{align}
  |DH|^2 + \text{Tr}\,|\mathcal D\Phi|^2
  &
  \supset
  - g \left(\frac{v}{2} \partial\varphi_H^+ + f \partial \varphi_\Phi^+ \right) W^-  + \text{h.c.} - g \frac{v}{2} \partial\varphi_H^0 A
  \ .
  \label{eq:Goldstone:eating}
\end{align}
A linear combination of $\varphi_H^{\pm}$ and $\varphi_{\Phi}^{\pm}$ is eaten by $W^{\pm}$ since both $\langle \Phi \rangle $ and $\langle H\rangle$ break $\SU{(2)}_V$ while $\varphi_H^0$ is only eaten by $A$ since the $\UU{(1)}_V$ subgroup is only broken by $\langle H \rangle$. We define the eaten Goldstone $\varphi_V$ and the orthogonal state $\varphi_A$ as
\begin{align}
  \varphi_V^\pm &=
  \frac{f \varphi_\Phi^\pm + (v/2)\varphi_H^\pm}{\sqrt{f^2 + (v/2)^2}}
  &
  \varphi_A^\pm &=
  \frac{f \varphi_H^\pm - (v/2)\varphi_\Phi^\pm}{\sqrt{f^2 + (v/2)^2}} \ . \label{eq:vectorandaxialstates}
\end{align}
$\varphi_A$ is the would-be Goldstone of the spontaneously broken axial symmetry .The Goldstone mode $\varphi_V^{\pm}$ is eaten by $W^{\pm}$ contributing to it's longitudinal polarization. When one ignores the trilinear term in \eqref{eq:V}, the $\varphi_A^{\pm}$ is massless at tree level. However, due to the broken axial symmetry, loops of the gauge bosons induce a radiative mass.

\subsection{vevs and Scalar Boson Masses}

Minimizing the potential \eqref{eq:V}, we find the vevs to be
\begin{align}
	f^2&=f_0^2+\frac{3 v^2}{\lambda}\left(\frac{\mu}{2 f}-\lambda''\right)
	&
	v^2&=v_0^2+\frac{3 f^2}{\lambda'}\left(\frac{\mu}{f}-\lambda''\right)\ . \label{eq:vevs2}
\end{align}
The radial modes $h$ and $\phi$ mix in this vacuum. In the basis $\begin{pmatrix} h,  & \phi \end{pmatrix}$, their mass matrix is given by
\begin{align}
	\mathcal{M}_H^2 &=
	\begin{pmatrix}
	\dfrac{\lambda' v^2}{3}&\lambda'' v f-\dfrac{\mu v}{2}\\
	\lambda'' v f-\dfrac{\mu v}{2} & \dfrac{\lambda f^2}{3} +\dfrac{\mu v^2}{4 f}
	\end{pmatrix}. \label{eq:MH2}
\end{align}
The eigenvalues of \eqref{eq:MH2} are
\begin{align}
m_{1,2}^2=\frac{1}{2}\text{Tr} \mathcal{M}_H^2 \mp \left|\frac{\left(\mathcal{M}_H^2\right)_{12}}{\sin 2\alpha}\right| \label{eq:m12}
\end{align}
where $\alpha$ is the angle that parameterizes the orthogonal transformation $\mathcal{O}_{\alpha}$ that diagonalizes \eqref{eq:MH2}. We define the transformation:
\begin{align}
  \begin{pmatrix}
  H_1
  \\
  H_2
  \end{pmatrix}
  &=
  \mathcal{O}_\alpha
  \begin{pmatrix}
  h
  \\
  \phi
  \end{pmatrix}
  &
  \mathcal{O}_\alpha &=
  \begin{pmatrix}
  \phantom{+}\cos\alpha & \sin\alpha
  \\
  -\sin\alpha & \cos\alpha
  \end{pmatrix} 
  \label{eq:Hrotation}
\end{align}
where the mixing angle $\alpha$ is related to the model parameters by
\begin{align}
	\tan 2\alpha =\frac{\mu v-2\lambda'' v f}{\lambda f^2/3 +\mu v^2 /4 f -\lambda' v^2/3}\ . \label{eq:tan2alpha}
\end{align}
We can expand \eqref{eq:m12} in the limit $v \ll f$ yielding the approximate eigenvalues
\begin{align}
m_1^2&= \frac{\lambda' v^2}{3}-\frac{3 v^2}{4 \lambda f^2}\left(\mu-2 f \lambda''\right)^2
+\mathcal{O}\left(\frac{v^4}{f^4}\right)
\label{eq:mpmorderv2_2:1}
\\
m_2^2 &= \frac{\lambda f^2}{3}+\frac{\mu v^2}{4 f}+\frac{3 v^2}{4 \lambda f^2}\left(\mu-2 f \lambda''\right)^2
+\mathcal{O}\left(\frac{v^4}{f^4}\right)
\ . \label{eq:mpmorderv2_2}
\end{align}
In this limit the masses of the radial modes form a hierarchy such that
\begin{align}
m_1^2 \sim \lambda' v^2 \ll m_2^2 \sim \lambda f^2 \ .
\end{align}

The trilinear term in \eqref{eq:V} causes the Goldstone modes to mix as well. As we did for the radial modes, we expand \eqref{eq:V} about the vacuum. In the basis $\begin{pmatrix} \varphi_\Phi^{\pm}  & \varphi_H^{\pm} \end{pmatrix}$ the Goldstone mass matrix is,
\begin{align}
\mathcal{M}_G^2
&=
\begin{pmatrix}
\phantom{+}\dfrac{\mu v^2}{4 f}& -\dfrac{\mu v}{2}\\
-\dfrac{\mu v}{2} & \phantom{+}\mu f
\end{pmatrix} \, .
\label{eq:MG}
\end{align}
It is easy to see that $\Det \mathcal{M}_G^2=0$, implying the existence of a zero eigenvalue corresponding to the Goldstone eaten by $W^{\pm}$. The other eigenvalue of \eqref{eq:MG} is simply $\tr \mathcal{M}_G^2$. As we did for the radial modes, we diagonalize $\mathcal{M}_G^2$ with the orthogonal transformation:
\begin{align}
  \begin{pmatrix}
  G^\pm
  \\
  \pi^\pm
  \end{pmatrix}
  &=
  \mathcal{O}_\beta
  \begin{pmatrix}
  \varphi_\Phi^\pm
  \\
  \varphi_H^\pm
  \end{pmatrix}
  &
  \mathcal{O}_\beta &=
  \begin{pmatrix}
  \phantom{+}\cos\beta & \sin\beta
  \\
  -\sin\beta & \cos\beta
  \end{pmatrix} 
  \label{eq:Goldstonerotation}
\end{align}
where $\pi^{\pm}$ is a massive \acro{PGB} and $G^{\pm}$ is the massless Goldstone eaten by $W^{\pm}$. The masses of these states are
\begin{align}
&m_G^2=0
&
&m_\pi^2=\mu f \left(1+\frac{v^2}{4f^2}\right) \,.
\end{align}
By requiring $\mathcal{O}_\beta$ diagonalize \eqref{eq:MG}, the mixing angle $\beta$ is found to be
\begin{align}
\tan\beta=\frac{v}{2f} \, . \label{eq:tanbeta}
\end{align}
The mass eigenstates defined by \eqref{eq:Goldstonerotation} are identical to the vector and axial states given by \eqref{eq:vectorandaxialstates}, confirming that $\SU{(2)}_A$ is indeed broken by both the gauging of $\SU{(2)}_V$ as well as the trilinear term in \eqref{eq:V}. We identify $\varphi_V^\pm=G^\pm$ and $\varphi_A^\pm=\pi^\pm$.

\section{Feynman Rules for Dark Sector States }
We outline the dark sector Feynman rules. Analogous to the interactions between the \acro{QCD} pion and Standard Model photon, the dark matter interactions with the dark photon are
\begin{align}
		\vcenter{
  		\hbox{\includegraphics[width=.2\textwidth]{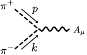}}
  		}
		&=
		\frac{ig}{4}\left(3-\cos2\beta\right)\left(p-k\right)_\mu
    \label{eq:pipiArule}
      \\
 	\vcenter{
  		\hbox{\includegraphics[width=.2\textwidth]{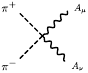}}
  		}
  &= 2 i g^2 \sin^2 \beta g_{\mu \nu}
  \, .
  \label{eq:pipiAArule}
    \end{align}
The Feynman rules for the dark matter interactions with the light radial mode are
\begin{align}
    \vcenter{
  		\hbox{\includegraphics[width=.2\textwidth]{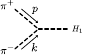}}
  		}
  =&\frac{i \left(p \cdot k\right)}{4 f}\left[\sin\left(\alpha-2\beta\right)+3\sin\left(\alpha+2\beta\right)\right]
\nonumber\\
	&
  -\frac{i}{f}\left(p\cdot k +m_\pi^2\right)\csc\beta \cos\left(\alpha-\beta\right)
  \label{eq:pipiH1rule}\\
    \vcenter{
  		\hbox{\includegraphics[width=.2\textwidth]{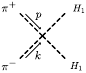}}
  		}
  =&
  -\frac{i}{8 f^2}\left(p \cdot k\right) \left[\cos 2\alpha\left(3\cos2\beta-7\right)+4\cos^2\alpha \csc^2\beta-5\cos 2\beta +1 \right]\nonumber\\[-20pt]
  &
  -\frac{i m_\pi^2}{2 f^2}\cos\alpha \cot\beta\left(\cos \alpha \cot\beta+4\sin\alpha \right)
  \, .
  \label{eq:pipiH1H1rule}
  \end{align}
  Similarly, the dark matter interactions with the heavy radial mode are
  \begin{align}
    \vcenter{
  		\hbox{\includegraphics[width=.2\textwidth]{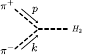}}
  		}
  =&\frac{i \left(p \cdot k\right)}{4 f}\left[\cos\left(\alpha-2\beta\right)+3\cos\left(\alpha+2\beta\right)\right]
\nonumber\\
&
  +\frac{i}{f}\left(p\cdot k +m_\pi^2\right)\csc\beta \sin\left(\alpha-\beta\right)
  \label{eq:pipiH2rule}\\
    \vcenter{
  		\hbox{\includegraphics[width=.2\textwidth]{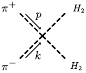}}
  		}
  =&
  -\frac{i \sin^2 \beta}{2f^2}\left(p \cdot k\right)\left(\sin^2\alpha \cot^4\beta+4 \cos^2\alpha\right)
  \nonumber\\
&
 -\frac{i m_\pi^2}{2 f^2}\sin \alpha \cot\beta \left(\sin\alpha \cot \beta-4\cos\alpha\right)
 \, .
  \label{eq:pipiH2H2rule}
\end{align}
The unbroken $\UU{(1)}_{H'}$ allows for the dark matter $\pi^\pm$ to interact with a $W^\pm$ of the opposite $\UU{(1)}_{H'}$ charge and a dark photon or radial mode. The Feynman rules are
\begin{align}
\vcenter{
      \hbox{\includegraphics[width=.2\textwidth]{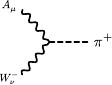}}
      }
&=\frac{g}{2} m_W \sin 2\beta\, g_{\mu  \nu } \label{eq:PipWmArule} 
\\
\vcenter{
      \hbox{\includegraphics[width=.2\textwidth]{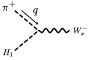}}
      }
&=g q_{\nu}\left(2 \sin \alpha \sin \beta -\cos\alpha \cos\beta\right) \label{eq:WpiH1rule}
\\
\vcenter{
      \hbox{\includegraphics[width=.2\textwidth]{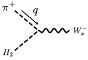}}
      }
&=g q_\nu \left(\sin\alpha \cos \beta+2\cos\alpha \sin \beta\right)
\,.
\label{eq:WpiH2rule}
\end{align}
These interactions mediate the tree-level decay of the $W^\pm$.

\section{Relic Abundance}
\label{sec:abundance}

\begin{figure}[t]
  \begin{center}
  \includegraphics[width=.22\textwidth]{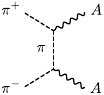}
  \;
  \includegraphics[width=.22\textwidth]{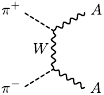}
  \;
  \includegraphics[width=.22\textwidth]{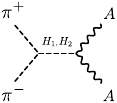}
  \;
  \includegraphics[width=.22\textwidth]{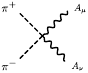}
\end{center}
\caption{Diagrams contributing to $\pi^+ \pi^- \to A A$ annihilation. Not shown: crossed ($u$-channel) diagrams and annihilation to scalars.}
\label{fig:pipiAAdiagrams}
\end{figure}

 Dark matter annihilation is dominated by $s$-wave processes in the low relative velocity limit. We consider the $\pi^+ \pi^- \to A A$, $H_1 H_1$, and $H_2H_2$ channels with all other possibilities being negligible at leading order in $v/f$. The diagrams contributing to the $\pi^+\pi^- \to A A$ process are shown schematically in Fig.~\ref{fig:pipiAAdiagrams}.
 To leading order in $v/f$ the annihilation cross sections are:
\begin{align}
\sigma v_{AA}
&=
\sigma_0 (m_A^2)
\left[\frac{1}{2}+\left(\frac{m_1^2}{4 m_A^2}-\frac{m_\pi^2}{2m_W^2}\frac{m_\pi^2+5m_W^2}{m_\pi^2+m_W^2}+X\right)^2\right]
\label{eq:sigmaAA}\\
\sigma v_{H_1 H_1}
&=
\sigma_0 (m_1^2)
\left[\frac{m_1^2}{2 m_A^2}+\frac{m_\pi^2}{2 m_W^2}+\frac{2m_\pi^2}{m_\pi^2+m_W^2}-X\right]^2
\label{eq:sigmaH1H1}\\
\sigma v _{H_2 H_2}
&=
\sigma_0(m_2^2) 
\left[
\frac{m_\pi^2}{4 m_W^2}+\frac{Y Z}{4m_W^2}-\frac{m_\pi^2}{m_W^2}Y^2
-\frac{ m_\pi^4}{m_W^2}\frac{ \left(3+Y\right)^2}{(m_2^2-4m_\pi^2)}+\frac{3 m_2^2}{16 m_W^2}\frac{ \left(2m_\pi^2-Z
\right)}{(m_2^2-4m_\pi^2)}\right]^2
\label{eq:sigmaH2H2}
\end{align}
where we have defined the functions
\begin{align}
\sigma_0 (m^2) &= \frac{\pi \alpha_X^2}{4 m_\pi^2}\sqrt{1-\frac{m^2}{m_\pi^2}}\\
X&=\frac{m_\pi^2}{2m_W^2}\frac{m_\pi^2-2 f^2 \lambda''}{m_2^2-4m_\pi^2}\left[1-\frac{2}{m_2^2}(m_\pi^2-2 f^2 \lambda'')\right]\\
Y&=\left(1-\frac{m_\pi^2-2 f^2 \lambda''}{m_2^2}\right)\\
Z&=\left(m_\pi^2-2 f^2 \lambda''\right)
\end{align}
as well as the dark fine structure constant
\begin{align}
\alpha_X=\frac{g^2}{4\pi}
\, .
\label{eq:darkfinestructure}
\end{align}

\begin{figure}[t]
  \begin{center}
  \includegraphics[width=\textwidth]{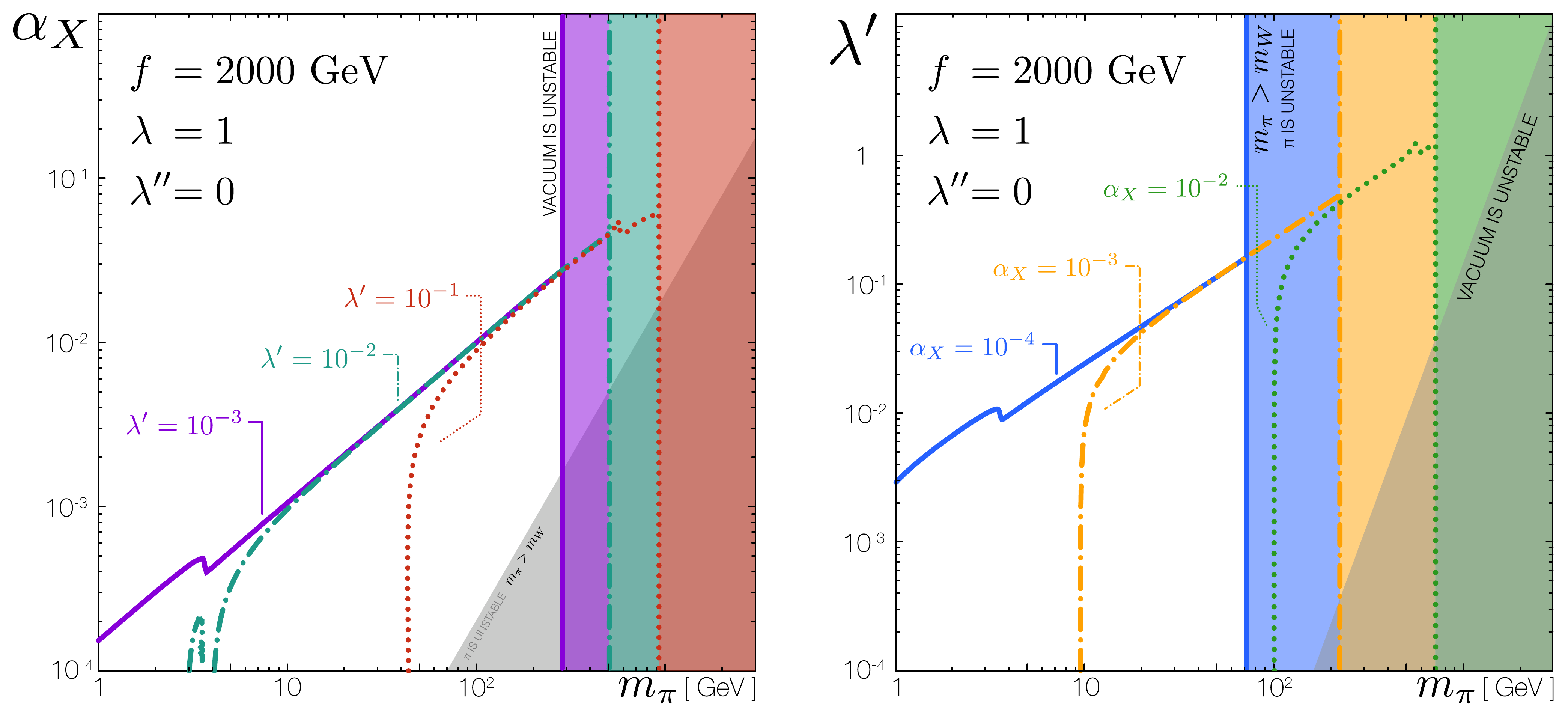}
\end{center}
\caption{The $\pi$ relic abundance as a function of the $\pi$ mass and dark fine structure constant/dark Higgs doublet coupling (Left/Right). The solid (\textsf{\color{Fuchsia}purple}/\textsf{\color{RoyalBlue}blue}), dash-dotted (\textsf{\color{JungleGreen}teal}/\textsf{\color{BurntOrange}orange}), and dotted (\textsf{\color{Mahogany}red}/\textsf{\color{OliveGreen}green}) curves represent when the $\pi$ saturate the dark matter relic abundance for $\lambda'=10^{-3}/\alpha_X=10^{-4}$, $10^{-2}/10^{-3}$, and $10^{-1}/10^{-2}$ respectively. The corresponding solid (\textsf{\color{Fuchsia}purple}/\textsf{\color{RoyalBlue}blue}), dash-dotted (\textsf{\color{JungleGreen}teal}/\textsf{\color{BurntOrange}orange}), and dotted (\textsf{\color{Mahogany}red}/\textsf{\color{OliveGreen}green}) vertical lines bound the regions in which the vacuum/$\pi$ are unstable. The shaded triangular regions denote where the $\pi$/vacuum is unstable.
}
\label{fig:relicabundance}
\end{figure}

We assume that the dark matter relic abundance is due to freeze-out in the early universe. The freeze-out temperature, $x_f=m_\pi/T_f$, and relic abundance, $\Omega h^2$, are~\cite{Kolb:1990vq}
\begin{align}
	x_f
  =&
  \ln \left(0.054 g_*^{-1/2}M_{Pl} \, m_\pi \langle\sigma v\rangle \right)
  -
  \frac{1}{2}\ln^2 \left(0.054 g_*^{-1/2} M_{Pl}\, m_\pi\langle\sigma v\rangle\right)
  \label{eq:xf}
  \\
	\Omega h^2 =& 2\times1.07 \times 10^9 \frac{x_f~\text{GeV}^{-1}}{\sqrt{g_*(x_f)}M_{Pl}\langle\sigma v\rangle} \label{eq:relicdensity} 
\end{align} 
where we have included an explicit factor of two in order to account for the fact that the dark matter may only annihilate with its anti-particle. The observed dark matter abundance is satisfied when $\Omega h^2 =0.12$~\cite{Tanabashi:2018oca, Steigman:2012nb}.

Fig.~\ref{fig:relicabundance} shows the values of $\alpha_X$ (left) and $\lambda'$ (right) which reproduce the observed relic abundance for a set of benchmark parameters. The curve indicating the observed relic abundance for a given $\lambda'/\alpha_X$ (Left/Right) becomes vertical when terms in the annihilation cross section independent of $\alpha_X /\lambda'$ (Left/Right) dominate. This imposes an effective lower bound on $m_\pi$ for a given set of benchmark parameters.

In this scenario we have implicitly assumed that the dark photon, $A$, is in equilibrium with the Standard Model thermal bath. We thus require
\begin{align}
 \Gamma_A \geq H(x_f\cong 20)
\end{align}
where $\Gamma_A$ is the dark photon decay width and $H(x_f)$ is the Hubble rate evaluated at the freeze-out temperature. The resulting constraint on the kinetic mixing parameter $\varepsilon$ in \eqref{eq:kinetic:mix:fermion:coupling} is~\cite{Pospelov:2007mp}:
\begin{align}
  \varepsilon^2 \left(\frac{m_A}{10~\text{MeV}}\right)\gtrsim 10^{-11}\left(\frac{m_\pi}{50~\text{GeV}}\right)^2 \ .
\end{align}
This assumption is not strictly necessary for a viable model. One may consider the case in which the dark sector is completely secluded, forming a thermal bath separate from the visible sector with a distinct initial temperature following reheating~\cite{Feng:2008mu}.  The thermal history of dark sectors with mediators has been studied more generally~\cite{Chu:2011be,Blennow:2013jba,Bernal:2015ova,Krnjaic:2017tio,Evans:2017kti, Dvorkin:2019zdi}. Another possible scenario is that UV dynamics generate an asymmetry in the $\pi$ abundance~\cite{Kaplan:2009ag,Turner:1987pp}. These situations are beyond the scope of this work and we leave them for future study.

\section{Self-Interactions}

Our dark sector automatically yields self-interactions among the $\pi$. Dark matter self-interactions were initially identified as a feature of dark sectors~\cite{Carlson:1992fn} and later observed to affect density profiles of dwarf galaxies~\cite{Spergel:1999mh,Dave:2000ar}. More recently, several small-scale structure anomalies have been connected to dark sector models with self-interactions~\cite{Feng:2009mn,Feng:2009hw,Buckley:2009in,Tulin:2013teo}. For a detailed review of the full parameter space of dark matter self-interactions see Ref.~\cite{Tulin:2017ara}.

The degree to which self-interactions affect dark matter halo densities depends on the scattering rate, $\sigma v \left( \rho_{\text{DM}}/m_{\text{DM}}\right)$. Because the dark matter relative velocity $v$ and density $\rho_{\text{DM}}$ are known for the astrophysical systems of interest, the relevant quantity is the ratio of the self-interaction cross section to the dark matter mass, $\sigma/m_{\text{DM}}$. Dwarf spheroidal galaxies with low relative velocities ($v\sim 10~$km/s) suffer from small-scale structure anomalies which may be alleviated in the presence of self-interactions~\cite{Tulin:2013teo,Kaplinghat:2015aga,Dave:2000ar}. Galaxy clusters, on the other hand, typically have larger relative velocities ($v\sim1500~$km/s) and similarly suffer from small-scale structure anomalies which may also be alleviated by self-interactions. The benchmark values for the ratio of the dark matter self-interaction cross section to its mass are
\begin{align}
&\left(\frac{\sigma}{m_{\text{DM}}}\right)_{\text{dwarf}} \sim 1~\frac{\text{cm}^2}{\text{g}}
&
&\left(\frac{\sigma}{m_{\text{DM}}}\right)_{\text{cluster}} \sim 0.1~\frac{\text{cm}^2}{\text{g}}
\, .
\label{eq:benchmark.sigma}
\end{align}
These seemingly inconsistent target cross sections may be achieved in tandem given the cross section has the appropriate velocity dependence.

The desired velocity dependence is achieved for non-relativistic scattering governed by a Yukawa potential. The dominant contribution to $\pi^\pm$ self interactions results from the exchange of dark photons, $A$, and yields a non-relativistic long-range scattering potential
\begin{align}
&V(r)
=
\pm\frac{\alpha_\pi}{r} e^{-m_A r}
&
\alpha_\pi
=
\frac{\alpha_X}{4}
\label{eq:VYukawa}
\end{align}
where the positive sign corresponds to particle-particle scattering and the negative sign to particle-antiparticle scattering. While the radial modes, $H_1$ and $H_2$, also contribute to self-scattering, the $\pi^+ \pi^- H_1$ vertex,  \eqref{eq:pipiH1rule}, is suppressed in the non-relativistic limit when $v \ll f$\footnote{Although the $\csc \beta$ term in \eqref{eq:pipiH1rule} diverges for $v\ll f$, for scattering $p \cdot k=-m_\pi^2$ and the coefficient of the divergent term vanishes exactly.}. Therefore we may ignore self-interactions mediated by $H_1$. On the other hand, $m_2 \gg m_A$ implies $H_2$ mediated self-interactions are sub-dominant compared to interactions mediated by the dark photon, $A$, due to the exponential suppression in \eqref{eq:VYukawa} and may be ignored as well.

\begin{figure}[ht]
  \includegraphics[width=\textwidth]{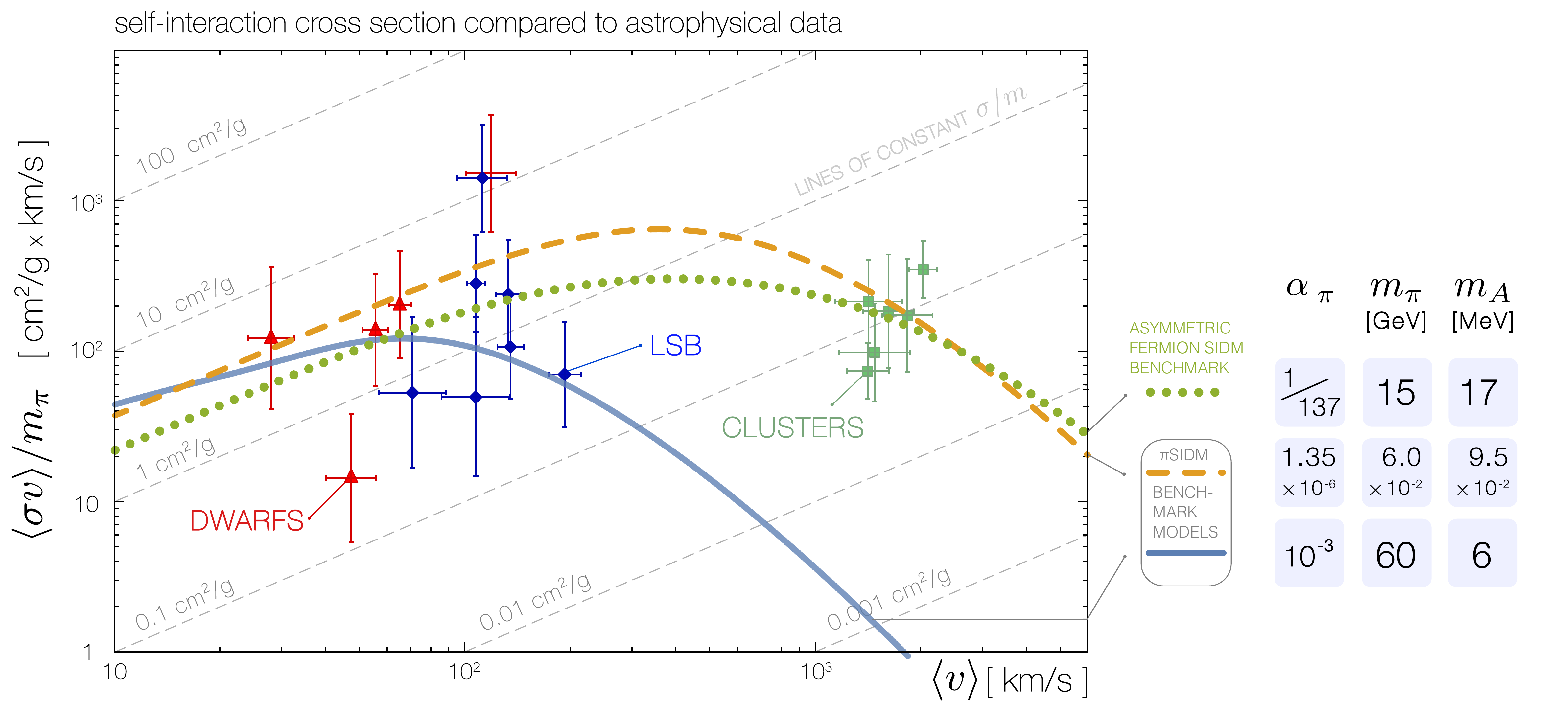}
  \caption{Numerical results for the dark matter self-interaction cross section in our model compared to cross sections for a set of dwarf galaxies, low surface brightness (\acro{LSB}) spiral galaxies, and galaxy clusters from Ref.~\cite{Kaplinghat:2015aga}. The solid/dashed (\textsf{\color{MidnightBlue}blue}/\textsf{\color{BurntOrange}orange}) curves corresponding to benchmarks with symmetric relic abundances, are compared to the dotted (\textsf{\color{OliveGreen}green}) curve corresponding to the benchmark model from Ref.~\cite{Kaplinghat:2015aga} with an asymmetric relic abundance. The benchmarks we present are identical to those found for spin-1 dark matter in Ref.~\cite{Chaffey:2019fec} with the replacements $m_W \to m_\pi$ and $\alpha_X \to \alpha_\pi$.}
  \label{fig:SIDMplot}
\end{figure}

The benchmark model of \acro{SIDM} consists of spin-1/2 dark matter with a mass $\sim 15~$GeV and a spin-1 mediator of mass $\sim 17~$MeV~\cite{Kaplinghat:2015aga}. The self-interaction potential is assumed to be purely repulsive, implying an asymmetry in the dark matter abundance. Cosmological constraints on dark matter annihilation in early universe typically favor models of asymmetric \acro{SIDM}, constraining  
\acro{SIDM} models with symmetric dark matter abundances to have sub-GeV scale masses~\cite{Huo:2017vef}. These constraints may be relaxed if we consider cluster scale density profile observations to be satisfied by some other mechanism, allowing for heavier dark matter masses.

Figure~\ref{fig:SIDMplot} compares two symmetric benchmark models in the $\pi$\acro{SIDM} framework to the asymmetric case studied in Ref.~\cite{Kaplinghat:2015aga}.
We numerically compute the self-interaction cross section following the methodology of Appendix B of Ref.~\cite{Chaffey:2019fec}, which is based on the procedure originally presented in Ref.\cite{Tulin:2013teo}.
The solid (\textsf{\color{MidnightBlue}blue}) curve only satisfies dwarf scale observations and corresponds to dark matter with mass $m_\pi = 60~\text{GeV}$, mediator mass $m_A = 6~\text{MeV}$, and coupling $\alpha_\pi = 10^{-3}$.
On the other hand, the dashed (\textsf{\color{BurntOrange}orange}) curve simultaneously satisfies both dwarf and cluster scale targets, corresponding to dark matter with mass $m_\pi = 60~\text{MeV}$, mediator mass $m_A = 95~\text{keV}$, and coupling $\alpha_\pi = 1.35 \times 10^{-6}$. We compare these benchmarks to the dotted (\textsf{\color{OliveGreen}green}) curve which reproduces the model from Ref.~\cite{Kaplinghat:2015aga} with $m_\pi = 15~\text{GeV}$, $m_A = 17~\text{MeV}$, and $\alpha_\pi = 1/137$. Because we consider contributions to the self-interaction cross section from both the repulsive and attractive potentials, our benchmarks are not necessarily unique due to the fact that an attractive potential displays resonant behavior~\cite{Tulin:2013teo}. In fact, these are the exact same benchmarks we present in Ref.~\cite{Chaffey:2019fec} with $m_W \to m_\pi$ and $\alpha_X \to \alpha_\pi$. In the standard freeze out scenario, our benchmark models may be fit to the observed relic abundance by tuning the parameters $\lambda$, $\lambda'$, and $f$. Ultimately, the cause of  dark matter halo density profile observations may be the result of contributions from baryonic feedback~\cite{Bullock:2017xww}. Therefore, we may interpret the data in Figure~\ref{fig:SIDMplot} as upper bounds on the self-interaction cross section.

\section{Portal Interactions}
\label{sec:Portal}

We consider a renormalizable vector portal interaction between our dark sector to the visible sector. Generally, one may also consider a scalar portal where the dark scalars $H$ and $\Phi$ couple to the Standard Model scalar sector through quartic interactions
\begin{align}
  \mathcal L \supset
  \lambda_{H\mathcal H} |H|^2|\mathcal H|^2  +
  \lambda_{\Phi\mathcal H} \left(\text{Tr}\,\Phi^2\right)|\mathcal H|^2 
  \label{eq:scalarportal}
\end{align}
where $\mathcal{H}$ is the Standard Model Higgs doublet. Models of \acro{PGBDM} which couple to the visible sector through a Higgs portal have been studied in Refs.~\cite{Gross:2017dan,Cline:2019okt,Karamitros:2019ewv,Okada:2020zxo}. In these models, the direct detection signature vanishes at zero momentum transfer as a result of a softly broken global symmetry.\footnote{In this context, softly broken refers to a symmetry group which is only broken by a mass term.} Because the axial symmetry group of our model is explicitly broken by a term trilinear in the fields, the direct detection cross section does not contain this feature. We consider the limit where the scalar portal is negligible compared to the vector portal in which the $\SU{(2)}_V$ field strength $F_{\mu\nu}=F_{\mu\nu}^a T^a$ and the adjoint triplet $\Phi$ may couple to the Standard model hypercharge field strength $\mathcal{B}^{\mu\nu}$ through the dimension-5 operator
\begin{align}
\frac{2}{\Lambda}\tr\left(\Phi F_{\mu \nu}\right)\mathcal{B}^{\mu \nu}
\end{align}
where $\Lambda$ is the scale of the \acro{UV} physics which generates this operator. The vev $\langle \Phi \rangle = f T^3$ induces kinetic mixing between dark photon and visible Standard model photon,
\begin{align}
\mathcal{L} \supset \frac{\varepsilon}{2 \cos\theta_W}F_{\mu \nu}\mathcal B^{\mu \nu}
    &\rightarrow  \frac{\varepsilon}{2} F_{\mu \nu}\mathcal F^{\mu \nu} \ ,
     \label{eq:Lkineticmixing}
\end{align}
where $\mathcal{F}^{\mu \nu}$ is the visible photon field strength. We do not consider mixing with the Z-boson as its contributions are negligible when the dark photon mass is much below the scale of electroweak symmetry breaking, $m_A \ll m_Z$.

The kinetic mixing given by \eqref{eq:Lkineticmixing} induces a coupling between the dark photon and the Standard Model electromagnetic current. This is consistent with the standard dark photon scenario, and may present signatures at present and future experiments~\cite{Alexander:2016aln,Battaglieri:2017aum}. The Feyman rule for the dark photon, $A$, and a fermion, $f$, with charge $Q_f$ is
\begin{align}
  \vcenter{
    \hbox{\includegraphics[width=.2\textwidth]{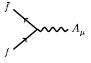}}
    }
  &=
  i \varepsilon e Q_f \gamma^\mu\ .
  \label{eq:kinetic:mix:fermion:coupling}
\end{align}
For bounds on the coupling $\varepsilon$ we refer to the reviews~\cite{Alexander:2016aln,Battaglieri:2017aum}, as our set up is identical to the standard dark photon.

To demonstrate the bounds on $\varepsilon$ from direct detection nucleon scattering experiments, we compute the scattering amplitude between dark matter, $\pi$, and a charged nucleon, $N$,
\begin{align}
  \vcenter{
    \hbox{\includegraphics[width=.2\textwidth]{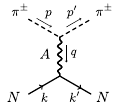}}
    }
  &=\pm\frac{i}{4} \frac{g \varepsilon e Q_N (3-\cos 2\beta)}{q^2-m_A^2}\bar{u}(k')\left(\slashed{p}+\slashed{p'}\right)u(k)\label{eq:Mnrelativistic}
  \ ,
\end{align}
which maps to a spin-independent operator $\mathcal O^{(\text{NR})}_1$ in the non-relativistic limit~\cite{Fan:2010gt,Fornengo:2011sz,DelNobile:2013sia,Dent:2015zpa}. Matching the notation of Ref.~\cite{Dent:2015zpa}, we identify
\begin{align}
&h_3 =\varepsilon e Q_q
&\text{and}
&&g_4 = \frac{g}{4}(3-\cos 2 \beta)
\, .
\end{align}
We define the effective coupling
\begin{align}
&c_1^N = -2 \frac{g_4 h_3^N}{m_A^2}
&\implies
&& c_p \equiv 
|c_1^N|
=\frac{\varepsilon e g}{2 m_A^2} (3-\cos 2\beta)
\simeq \frac{\varepsilon e g}{m_A^2}
\end{align}
where we have assumed the limit $\tan \beta = v/2 f\ll 1$. Due to the conservation of the electromagnetic charge, the effective coupling $h_3^N$ for a nucleon, $N$, is simply proportional to the charge of the nucleon.

\begin{figure}[ht]
  \includegraphics[width=\textwidth]{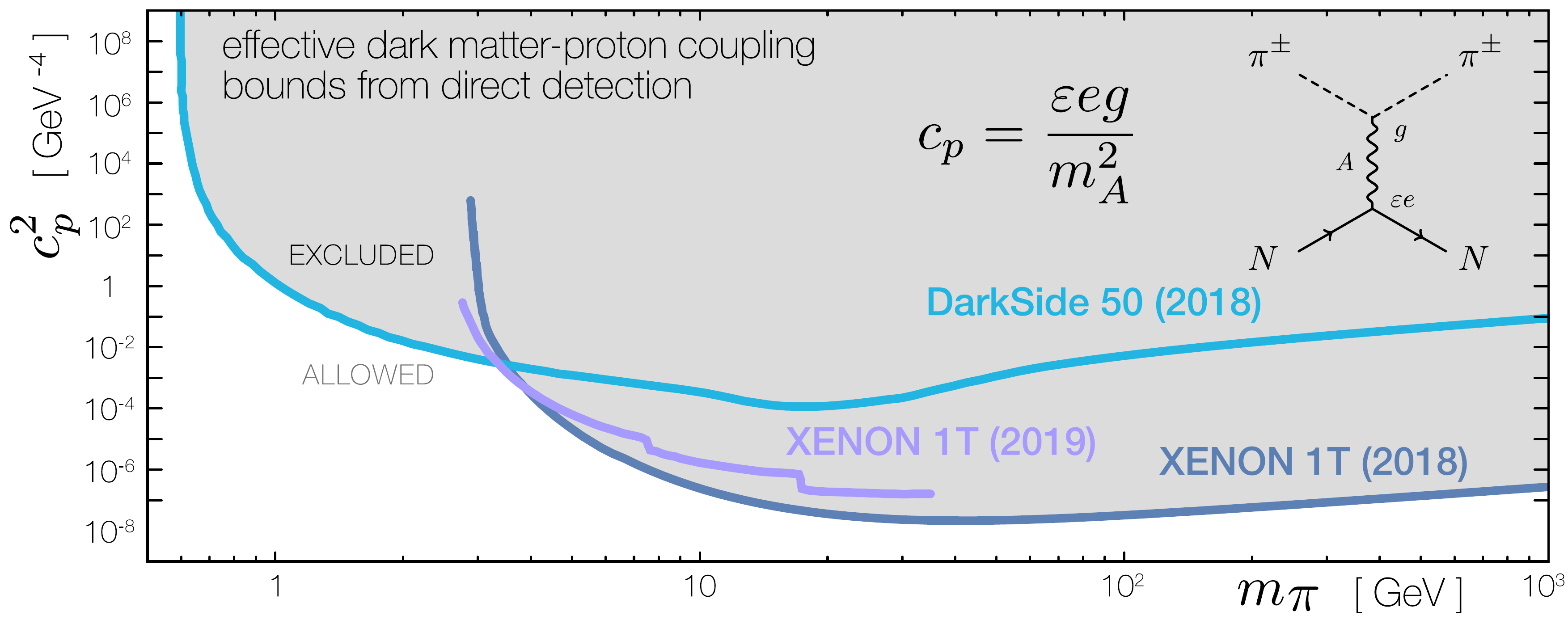}
  \caption{Constraints on the effective dark matter--proton coupling, $c_p^2$, from direct detection experiments  \acro{XENON~1T}~\cite{Aprile:2018dbl,XENON:2019gfn} and DarkSide 50~\cite{Agnes:2018fwg}.}
  \label{fig:DDbounds}
\end{figure}

The effective coupling $c_p$ is constrained by spin-independent dark matter-nucleon scattering from searches such as \acro{XENON~1T}~\cite{Aprile:2018dbl,XENON:2019gfn} and DarkSide 50~\cite{Agnes:2018fwg}. We compare the bounds from these searches to the effective coupling $c_p$ in Fig.~\ref{fig:DDbounds}. For a fixed mediator mass $m_A$ and dark gauge coupling $g$, the spin-independent dark matter-nucleon cross section sets an upper bound on the kinetic mixing parameter, $\varepsilon$.  While beyond the scope of this study, for values of $\varepsilon$ so small that the dark sector is effectively decoupled from the visible sector, one may produce thermal histories distinct from the thermal freeze-out scenario~\cite{Chu:2011be,Blennow:2013jba,Krnjaic:2017tio,Evans:2017kti, Dvorkin:2019zdi}.

\section{Conclusion}
We present a model of pseudo-Goldstone boson dark matter and dark photon mediator. The dark matter mass is finite due to the explicit breaking of the axial subgroup of an $\SU{(2)} \times \SU{(2)}$ symmetry. Spontaneous symmetry breaking sets the scale of the dark matter and mediator masses, realizing a rich spectrum of states. A residual global $\UU{(1)}$ stabilizes the psuedo-Goldstone states which are assumed to  be lightest in the spectrum charged under this symmetry.
We find that the \acro{PGB} states may saturate the observed dark matter relic abundance in the standard thermal freeze-out scenario. For certain benchmark models some small-scale structure anomalies may be resolved by dark matter self-interactions, however the requirement that the dark matter be a thermal relic makes fitting the self-interaction cross section to observations on dwarf galaxy and cluster scales simultaneously difficult. We leave a precise fit of the self-interaction cross section to observed small-scale structure anomalies for future work.
We present direct detection bounds on our \acro{PGBDM} which is assumed to primarily interact with the visible sector through a vector portal. In general a Higgs portal may be considered. However, such interactions introduce further mixing between the radial modes and thus are beyond the scope of this study.

The model presented in this work offers several avenues for further study. One may consider inelastic scattering off of nucleons, in which a $\pi^{\pm}$ up-scatters off of a nucleon producing a $W^\pm$. This model may be understood as the phase of the model of spin-1 self-interacting dark matter we present in Ref.~\cite{Chaffey:2019fec}, where $m_W >m_\pi$. Thus, a natural extension of these scenarios is to consider multi-component dark matter in which the observed dark matter abundance consists of a combination of $\pi^\pm$ and $W^\pm$. Such models describe inelastic dark matter which may admit novel phenomenology.
Another exciting possibility is to restore the Higgs portal interaction with the visible sector.

\section*{Acknowledgments}
We thank Flip Tanedo for insightful discussions. We thank Meghan Neureither and Lemon Neureither-Chaffey for their love and support. We are supported by the DOE grant \acro{DE}-\acro{SC}/0008541.

\bibliographystyle{utcaps} 	
\bibliography{PGDM.bib}

\providecommand{\href}[2]{#2}\begingroup\raggedright\begin{thebibliography}{10}

\bibitem{Ade:2015xua}
{\bfseries Planck} Collaboration, P.~A.~R. Ade {\em et~al.}, ``{Planck 2015
  results. XIII. Cosmological parameters},''
  \href{http://dx.doi.org/10.1051/0004-6361/201525830}{{\em Astron. Astrophys.}
  {\bfseries 594} (2016) A13},
\href{http://arxiv.org/abs/1502.01589}{{\ttfamily arXiv:1502.01589
  [astro-ph.CO]}}.

\bibitem{Aghanim:2018eyx}
{\bfseries Planck} Collaboration, N.~Aghanim {\em et~al.}, ``{Planck 2018
  results. VI. Cosmological parameters},''
\href{http://arxiv.org/abs/1807.06209}{{\ttfamily arXiv:1807.06209
  [astro-ph.CO]}}.

\bibitem{Tan:2016zwf}
{\bfseries PandaX-II} Collaboration, A.~Tan {\em et~al.}, ``{Dark Matter
  Results from First 98.7 Days of Data from the PandaX-II Experiment},''
  \href{http://dx.doi.org/10.1103/PhysRevLett.117.121303}{{\em Phys. Rev.
  Lett.} {\bfseries 117} no.~12, (2016) 121303},
\href{http://arxiv.org/abs/1607.07400}{{\ttfamily arXiv:1607.07400 [hep-ex]}}.

\bibitem{Akerib:2015rjg}
{\bfseries LUX} Collaboration, D.~S. Akerib {\em et~al.}, ``{Improved Limits on
  Scattering of Weakly Interacting Massive Particles from Reanalysis of 2013
  LUX Data},'' \href{http://dx.doi.org/10.1103/PhysRevLett.116.161301}{{\em
  Phys. Rev. Lett.} {\bfseries 116} no.~16, (2016) 161301},
\href{http://arxiv.org/abs/1512.03506}{{\ttfamily arXiv:1512.03506
  [astro-ph.CO]}}.

\bibitem{Akerib:2016vxi}
{\bfseries LUX} Collaboration, D.~S. Akerib {\em et~al.}, ``{Results from a
  search for dark matter in the complete LUX exposure},''
  \href{http://dx.doi.org/10.1103/PhysRevLett.118.021303}{{\em Phys. Rev.
  Lett.} {\bfseries 118} no.~2, (2017) 021303},
\href{http://arxiv.org/abs/1608.07648}{{\ttfamily arXiv:1608.07648
  [astro-ph.CO]}}.

\bibitem{Agnese:2015nto}
{\bfseries SuperCDMS} Collaboration, R.~Agnese {\em et~al.}, ``{New Results
  from the Search for Low-Mass Weakly Interacting Massive Particles with the
  CDMS Low Ionization Threshold Experiment},''
  \href{http://dx.doi.org/10.1103/PhysRevLett.116.071301}{{\em Phys. Rev.
  Lett.} {\bfseries 116} no.~7, (2016) 071301},
\href{http://arxiv.org/abs/1509.02448}{{\ttfamily arXiv:1509.02448
  [astro-ph.CO]}}.

\bibitem{Hooper:2012sr}
D.~Hooper, C.~Kelso, and F.~S. Queiroz, ``{Stringent and Robust Constraints on
  the Dark Matter Annihilation Cross Section From the Region of the Galactic
  Center},'' \href{http://dx.doi.org/10.1016/j.astropartphys.2013.04.007}{{\em
  Astropart. Phys.} {\bfseries 46} (2013) 55--70},
\href{http://arxiv.org/abs/1209.3015}{{\ttfamily arXiv:1209.3015
  [astro-ph.HE]}}.

\bibitem{Ackermann:2015zua}
{\bfseries Fermi-LAT} Collaboration, M.~Ackermann {\em et~al.}, ``{Searching
  for Dark Matter Annihilation from Milky Way Dwarf Spheroidal Galaxies with
  Six Years of Fermi Large Area Telescope Data},''
  \href{http://dx.doi.org/10.1103/PhysRevLett.115.231301}{{\em Phys. Rev.
  Lett.} {\bfseries 115} no.~23, (2015) 231301},
\href{http://arxiv.org/abs/1503.02641}{{\ttfamily arXiv:1503.02641
  [astro-ph.HE]}}.

\bibitem{Bergstrom:2013jra}
L.~Bergstrom, T.~Bringmann, I.~Cholis, D.~Hooper, and C.~Weniger, ``{New Limits
  on Dark Matter Annihilation from AMS Cosmic Ray Positron Data},''
  \href{http://dx.doi.org/10.1103/PhysRevLett.111.171101}{{\em Phys. Rev.
  Lett.} {\bfseries 111} (2013) 171101},
\href{http://arxiv.org/abs/1306.3983}{{\ttfamily arXiv:1306.3983
  [astro-ph.HE]}}.

\bibitem{Cirelli:2013hv}
M.~Cirelli and G.~Giesen, ``{Antiprotons from Dark Matter: Current constraints
  and future sensitivities},''
  \href{http://dx.doi.org/10.1088/1475-7516/2013/04/015}{{\em JCAP} {\bfseries
  1304} (2013) 015},
\href{http://arxiv.org/abs/1301.7079}{{\ttfamily arXiv:1301.7079 [hep-ph]}}.

\bibitem{Pospelov:2008zw}
M.~Pospelov, ``{Secluded U(1) Below the Weak Scale},''
  \href{http://dx.doi.org/10.1103/PhysRevD.80.095002}{{\em Phys. Rev.}
  {\bfseries D80} (2009) 095002},
\href{http://arxiv.org/abs/0811.1030}{{\ttfamily arXiv:0811.1030 [hep-ph]}}.

\bibitem{Pospelov:2008jd}
M.~Pospelov and A.~Ritz, ``{Astrophysical Signatures of Secluded Dark
  Matter},'' \href{http://dx.doi.org/10.1016/j.physletb.2008.12.012}{{\em Phys.
  Lett.} {\bfseries B671} (2009) 391--397},
\href{http://arxiv.org/abs/0810.1502}{{\ttfamily arXiv:0810.1502 [hep-ph]}}.

\bibitem{Pospelov:2007mp}
M.~Pospelov, A.~Ritz, and M.~B. Voloshin, ``{Secluded WIMP Dark Matter},''
  \href{http://dx.doi.org/10.1016/j.physletb.2008.02.052}{{\em Phys. Lett.}
  {\bfseries B662} (2008) 53--61},
\href{http://arxiv.org/abs/0711.4866}{{\ttfamily arXiv:0711.4866 [hep-ph]}}.

\bibitem{Essig:2013lka}
R.~Essig {\em et~al.}, ``{Working Group Report: New Light Weakly Coupled
  Particles},'' in {\em {Proceedings, 2013 Community Summer Study on the Future
  of U.S. Particle Physics: Snowmass on the Mississippi (CSS2013): Minneapolis,
  MN, USA, July 29-August 6, 2013}}.
\newblock 2013.
\newblock \href{http://arxiv.org/abs/1311.0029}{{\ttfamily arXiv:1311.0029
  [hep-ph]}}.
\newblock
\url{http://www.slac.stanford.edu/econf/C1307292/docs/IntensityFrontier/NewLight-17.pdf}.
\newblock

\bibitem{Alexander:2016aln}
J.~Alexander {\em et~al.}, ``{Dark Sectors 2016 Workshop: Community Report},''
\newblock 2016.
\newblock \href{http://arxiv.org/abs/1608.08632}{{\ttfamily arXiv:1608.08632
  [hep-ph]}}.
\newblock
\url{http://inspirehep.net/record/1484628/files/arXiv:1608.08632.pdf}.
\newblock

\bibitem{Battaglieri:2017aum}
M.~Battaglieri {\em et~al.}, ``{Us Cosmic Visions: New Ideas in Dark Matter
  2017: Community Report},'' in {\em {U.S. Cosmic Visions: New Ideas in Dark
  Matter College Park, Md, Usa, March 23-25, 2017}}.
\newblock 2017.
\newblock \href{http://arxiv.org/abs/1707.04591}{{\ttfamily arXiv:1707.04591
  [hep-ph]}}.
\newblock
\url{http://lss.fnal.gov/archive/2017/conf/fermilab-conf-17-282-ae-ppd-t.pdf}.
\newblock

\bibitem{Tulin:2017ara}
S.~Tulin and H.-B. Yu, ``{Dark Matter Self-interactions and Small Scale
  Structure},'' \href{http://dx.doi.org/10.1016/j.physrep.2017.11.004}{{\em
  Phys. Rept.} {\bfseries 730} (2018) 1--57},
\href{http://arxiv.org/abs/1705.02358}{{\ttfamily arXiv:1705.02358 [hep-ph]}}.

\bibitem{SILVEIRA1985136}
V.~Silveira and A.~Zee, ``Scalar Phantoms,''
  \href{http://dx.doi.org/https://doi.org/10.1016/0370-2693(85)90624-0}{{\em
  Physics Letters B} {\bfseries 161} no.~1, (1985) 136--140}.
  \url{https://www.sciencedirect.com/science/article/pii/0370269385906240}.

\bibitem{McDonald:1993ex}
J.~McDonald, ``{Gauge singlet scalars as cold dark matter},''
  \href{http://dx.doi.org/10.1103/PhysRevD.50.3637}{{\em Phys. Rev. D}
  {\bfseries 50} (1994) 3637--3649},
  \href{http://arxiv.org/abs/hep-ph/0702143}{{\ttfamily arXiv:hep-ph/0702143}}.

\bibitem{Burgess_2001}
C.~Burgess, M.~Pospelov, and T.~ter Veldhuis, ``The Minimal Model of
  nonbaryonic dark matter: a singlet scalar,''
  \href{http://dx.doi.org/10.1016/s0550-3213(01)00513-2}{{\em Nuclear Physics
  B} {\bfseries 619} no.~1-3, (Dec, 2001) 709–728}.
  \url{http://dx.doi.org/10.1016/S0550-3213(01)00513-2}.

\bibitem{Cohen_2012}
T.~Cohen, J.~Kearney, A.~Pierce, and D.~Tucker-Smith, ``Singlet-doublet dark
  matter,'' \href{http://dx.doi.org/10.1103/physrevd.85.075003}{{\em Physical
  Review D} {\bfseries 85} no.~7, (Apr, 2012) }.
  \url{http://dx.doi.org/10.1103/PhysRevD.85.075003}.

\bibitem{Ma_2006}
E.~Ma, ``Verifiable radiative seesaw mechanism of neutrino mass and dark
  matter,'' \href{http://dx.doi.org/10.1103/physrevd.73.077301}{{\em Physical
  Review D} {\bfseries 73} no.~7, (Apr, 2006) }.
  \url{http://dx.doi.org/10.1103/PhysRevD.73.077301}.

\bibitem{Honorez_2007}
L.~L. Honorez, E.~Nezri, J.~F. Oliver, and M.~H.~G. Tytgat, ``The inert doublet
  model: an archetype for dark matter,''
  \href{http://dx.doi.org/10.1088/1475-7516/2007/02/028}{{\em Journal of
  Cosmology and Astroparticle Physics} {\bfseries 2007} no.~02, (Feb, 2007)
  028–028}. \url{http://dx.doi.org/10.1088/1475-7516/2007/02/028}.

\bibitem{Escudero_2016}
M.~Escudero, A.~Berlin, D.~Hooper, and M.-X. Lin, ``Toward (finally!) ruling
  out Z and Higgs mediated dark matter models,''
  \href{http://dx.doi.org/10.1088/1475-7516/2016/12/029}{{\em Journal of
  Cosmology and Astroparticle Physics} {\bfseries 2016} no.~12, (Dec, 2016)
  029–029}. \url{http://dx.doi.org/10.1088/1475-7516/2016/12/029}.

\bibitem{B_hm_2004}
C.~Bœhm and P.~Fayet, ``Scalar dark matter candidates,''
  \href{http://dx.doi.org/10.1016/j.nuclphysb.2004.01.015}{{\em Nuclear Physics
  B} {\bfseries 683} no.~1-2, (Apr, 2004) 219–263}.
  \url{http://dx.doi.org/10.1016/j.nuclphysb.2004.01.015}.

\bibitem{Boehm:2020wbt}
C.~Boehm, X.~Chu, J.-L. Kuo, and J.~Pradler, ``{Scalar dark matter candidates
  revisited},'' \href{http://dx.doi.org/10.1103/PhysRevD.103.075005}{{\em Phys.
  Rev. D} {\bfseries 103} no.~7, (2021) 075005},
  \href{http://arxiv.org/abs/2010.02954}{{\ttfamily arXiv:2010.02954
  [hep-ph]}}.

\bibitem{Barger_2009}
V.~Barger, P.~Langacker, M.~McCaskey, M.~Ramsey-Musolf, and G.~Shaughnessy,
  ``Complex singlet extension of the standard model,''
  \href{http://dx.doi.org/10.1103/physrevd.79.015018}{{\em Physical Review D}
  {\bfseries 79} no.~1, (Jan, 2009) }.
  \url{http://dx.doi.org/10.1103/PhysRevD.79.015018}.

\bibitem{Frigerio:2012uc}
M.~Frigerio, A.~Pomarol, F.~Riva, and A.~Urbano, ``{Composite Scalar Dark
  Matter},'' \href{http://dx.doi.org/10.1007/JHEP07(2012)015}{{\em JHEP}
  {\bfseries 07} (2012) 015}, \href{http://arxiv.org/abs/1204.2808}{{\ttfamily
  arXiv:1204.2808 [hep-ph]}}.

\bibitem{Gross:2017dan}
C.~Gross, O.~Lebedev, and T.~Toma, ``{Cancellation Mechanism for
  Dark-Matter–Nucleon Interaction},''
  \href{http://dx.doi.org/10.1103/PhysRevLett.119.191801}{{\em Phys. Rev.
  Lett.} {\bfseries 119} no.~19, (2017) 191801},
\href{http://arxiv.org/abs/1708.02253}{{\ttfamily arXiv:1708.02253 [hep-ph]}}.

\bibitem{Cline:2019okt}
J.~M. Cline and T.~Toma, ``{Pseudo-Goldstone dark matter confronts cosmic ray
  and collider anomalies},''
\href{http://arxiv.org/abs/1906.02175}{{\ttfamily arXiv:1906.02175 [hep-ph]}}.

\bibitem{Karamitros:2019ewv}
D.~Karamitros, ``{Pseudo Nambu-Goldstone Dark Matter: Examples of Vanishing
  Direct Detection Cross Section},''
  \href{http://dx.doi.org/10.1103/PhysRevD.99.095036}{{\em Phys. Rev.}
  {\bfseries D99} no.~9, (2019) 095036},
\href{http://arxiv.org/abs/1901.09751}{{\ttfamily arXiv:1901.09751 [hep-ph]}}.

\bibitem{Okada:2020zxo}
N.~Okada, D.~Raut, and Q.~Shafi, ``{Pseudo-Goldstone dark matter in a gauged
  $B-L$ extended standard model},''
  \href{http://dx.doi.org/10.1103/PhysRevD.103.055024}{{\em Phys. Rev. D}
  {\bfseries 103} no.~5, (2021) 055024},
  \href{http://arxiv.org/abs/2001.05910}{{\ttfamily arXiv:2001.05910
  [hep-ph]}}.

\bibitem{Chaffey:2019fec}
I.~Chaffey and P.~Tanedo, ``{Vector Self-Interacting Dark Matter},''
\href{http://arxiv.org/abs/1907.10217}{{\ttfamily arXiv:1907.10217 [hep-ph]}}.

\bibitem{Holdom:1985ag}
B.~Holdom, ``{Two U(1)'s and Epsilon Charge Shifts},''
\href{http://dx.doi.org/10.1016/0370-2693(86)91377-8}{{\em Phys. Lett.}
  {\bfseries 166B} (1986) 196--198}.

\bibitem{GALISON1984279}
P.~Galison and A.~Manohar, ``Two Z's or not two Z's?,''
  \href{http://dx.doi.org/https://doi.org/10.1016/0370-2693(84)91161-4}{{\em
  Physics Letters B} {\bfseries 136} no.~4, (1984) 279 -- 283}.
  \url{http://www.sciencedirect.com/science/article/pii/0370269384911614}.

\bibitem{Tsai:2020vpi}
Y.-D. Tsai, R.~McGehee, and H.~Murayama, ``{Resonant Self-Interacting Dark
  Matter from Dark QCD},'' \href{http://arxiv.org/abs/2008.08608}{{\ttfamily
  arXiv:2008.08608 [hep-ph]}}.

\bibitem{Hall:2019rld}
E.~Hall, T.~Konstandin, R.~McGehee, and H.~Murayama, ``{Asymmetric Matters from
  a Dark First-Order Phase Transition},''
  \href{http://arxiv.org/abs/1911.12342}{{\ttfamily arXiv:1911.12342
  [hep-ph]}}.

\bibitem{Hall:2021zsk}
E.~Hall, R.~McGehee, H.~Murayama, and B.~Suter, ``{Asymmetric Dark Matter May
  Not Be Light},'' \href{http://arxiv.org/abs/2107.03398}{{\ttfamily
  arXiv:2107.03398 [hep-ph]}}.

\bibitem{PhysRev.177.2247}
C.~G. Callan, S.~Coleman, J.~Wess, and B.~Zumino, ``Structure of
  Phenomenological Lagrangians. II,''
  \href{http://dx.doi.org/10.1103/PhysRev.177.2247}{{\em Phys. Rev.} {\bfseries
  177} (Jan, 1969) 2247--2250}.
  \url{https://link.aps.org/doi/10.1103/PhysRev.177.2247}.

\bibitem{Kolb:1990vq}
E.~W. Kolb and M.~S. Turner, ``{The Early Universe},''
{\em Front. Phys.} {\bfseries 69} (1990) 1--547.

\bibitem{Tanabashi:2018oca}
{\bfseries Particle Data Group} Collaboration, M.~Tanabashi {\em et~al.},
  ``{Review of Particle Physics},''
\href{http://dx.doi.org/10.1103/PhysRevD.98.030001}{{\em Phys. Rev.} {\bfseries
  D98} no.~3, (2018) 030001}.

\bibitem{Steigman:2012nb}
G.~Steigman, B.~Dasgupta, and J.~F. Beacom, ``{Precise Relic WIMP Abundance and
  Its Impact on Searches for Dark Matter Annihilation},''
  \href{http://dx.doi.org/10.1103/PhysRevD.86.023506}{{\em Phys. Rev.}
  {\bfseries D86} (2012) 023506},
\href{http://arxiv.org/abs/1204.3622}{{\ttfamily arXiv:1204.3622 [hep-ph]}}.

\bibitem{Feng:2008mu}
J.~L. Feng, H.~Tu, and H.-B. Yu, ``{Thermal Relics in Hidden Sectors},''
  \href{http://dx.doi.org/10.1088/1475-7516/2008/10/043}{{\em JCAP} {\bfseries
  0810} (2008) 043},
\href{http://arxiv.org/abs/0808.2318}{{\ttfamily arXiv:0808.2318 [hep-ph]}}.

\bibitem{Chu:2011be}
X.~Chu, T.~Hambye, and M.~H.~G. Tytgat, ``{The Four Basic Ways of Creating Dark
  Matter Through a Portal},''
  \href{http://dx.doi.org/10.1088/1475-7516/2012/05/034}{{\em JCAP} {\bfseries
  1205} (2012) 034},
\href{http://arxiv.org/abs/1112.0493}{{\ttfamily arXiv:1112.0493 [hep-ph]}}.

\bibitem{Blennow:2013jba}
M.~Blennow, E.~Fernandez-Mart{\'\i ne}z, and B.~Zaldivar, ``{Freeze-In Through
  Portals},'' \href{http://dx.doi.org/10.1088/1475-7516/2014/01/003}{{\em JCAP}
  {\bfseries 1401} (2014) 003},
\href{http://arxiv.org/abs/1309.7348}{{\ttfamily arXiv:1309.7348 [hep-ph]}}.

\bibitem{Bernal:2015ova}
N.~Bernal, X.~Chu, C.~Garcia-Cely, T.~Hambye, and B.~Zaldivar, ``{Production
  Regimes for Self-Interacting Dark Matter},''
  \href{http://dx.doi.org/10.1088/1475-7516/2016/03/018}{{\em JCAP} {\bfseries
  1603} no.~03, (2016) 018},
\href{http://arxiv.org/abs/1510.08063}{{\ttfamily arXiv:1510.08063 [hep-ph]}}.

\bibitem{Krnjaic:2017tio}
G.~Krnjaic, ``{Freezing In, Heating Up, and Freezing Out: Predictive Nonthermal
  Dark Matter and Low-Mass Direct Detection},''
  \href{http://dx.doi.org/10.1007/JHEP10(2018)136}{{\em JHEP} {\bfseries 10}
  (2018) 136},
\href{http://arxiv.org/abs/1711.11038}{{\ttfamily arXiv:1711.11038 [hep-ph]}}.

\bibitem{Evans:2017kti}
J.~A. Evans, S.~Gori, and J.~Shelton, ``{Looking for the WIMP Next Door},''
  \href{http://dx.doi.org/10.1007/JHEP02(2018)100}{{\em JHEP} {\bfseries 02}
  (2018) 100},
\href{http://arxiv.org/abs/1712.03974}{{\ttfamily arXiv:1712.03974 [hep-ph]}}.

\bibitem{Dvorkin:2019zdi}
C.~Dvorkin, T.~Lin, and K.~Schutz, ``{Making Dark Matter Out of Light:
  Freeze-In from Plasma Effects},''
  \href{http://dx.doi.org/10.1103/PhysRevD.99.115009}{{\em Phys. Rev.}
  {\bfseries D99} no.~11, (2019) 115009},
\href{http://arxiv.org/abs/1902.08623}{{\ttfamily arXiv:1902.08623 [hep-ph]}}.

\bibitem{Kaplan:2009ag}
D.~E. Kaplan, M.~A. Luty, and K.~M. Zurek, ``{Asymmetric Dark Matter},''
  \href{http://dx.doi.org/10.1103/PhysRevD.79.115016}{{\em Phys. Rev.}
  {\bfseries D79} (2009) 115016},
\href{http://arxiv.org/abs/0901.4117}{{\ttfamily arXiv:0901.4117 [hep-ph]}}.

\bibitem{Turner:1987pp}
M.~S. Turner and B.~J. Carr, ``{Why Should Baryons and Exotic Relic Particles
  Have Comparable Densities?},''
\href{http://dx.doi.org/10.1142/S0217732387000021}{{\em Mod. Phys. Lett.}
  {\bfseries A2} (1987) 1--7}.

\bibitem{Carlson:1992fn}
E.~D. Carlson, M.~E. Machacek, and L.~J. Hall, ``{Self-Interacting Dark
  Matter},''
\href{http://dx.doi.org/10.1086/171833}{{\em Astrophys. J.} {\bfseries 398}
  (1992) 43--52}.

\bibitem{Spergel:1999mh}
D.~N. Spergel and P.~J. Steinhardt, ``{Observational Evidence for
  Selfinteracting Cold Dark Matter},''
  \href{http://dx.doi.org/10.1103/PhysRevLett.84.3760}{{\em Phys. Rev. Lett.}
  {\bfseries 84} (2000) 3760--3763},
\href{http://arxiv.org/abs/astro-ph/9909386}{{\ttfamily arXiv:astro-ph/9909386
  [astro-ph]}}.

\bibitem{Dave:2000ar}
R.~Dave, D.~N. Spergel, P.~J. Steinhardt, and B.~D. Wandelt, ``{Halo Properties
  in Cosmological Simulations of Selfinteracting Cold Dark Matter},''
  \href{http://dx.doi.org/10.1086/318417}{{\em Astrophys. J.} {\bfseries 547}
  (2001) 574--589},
\href{http://arxiv.org/abs/astro-ph/0006218}{{\ttfamily arXiv:astro-ph/0006218
  [astro-ph]}}.

\bibitem{Feng:2009mn}
J.~L. Feng, M.~Kaplinghat, H.~Tu, and H.-B. Yu, ``{Hidden Charged Dark
  Matter},'' \href{http://dx.doi.org/10.1088/1475-7516/2009/07/004}{{\em JCAP}
  {\bfseries 0907} (2009) 004},
\href{http://arxiv.org/abs/0905.3039}{{\ttfamily arXiv:0905.3039 [hep-ph]}}.

\bibitem{Feng:2009hw}
J.~L. Feng, M.~Kaplinghat, and H.-B. Yu, ``{Halo Shape and Relic Density
  Exclusions of Sommerfeld-Enhanced Dark Matter Explanations of Cosmic Ray
  Excesses},'' \href{http://dx.doi.org/10.1103/PhysRevLett.104.151301}{{\em
  Phys. Rev. Lett.} {\bfseries 104} (2010) 151301},
\href{http://arxiv.org/abs/0911.0422}{{\ttfamily arXiv:0911.0422 [hep-ph]}}.

\bibitem{Buckley:2009in}
M.~R. Buckley and P.~J. Fox, ``{Dark Matter Self-Interactions and Light Force
  Carriers},'' \href{http://dx.doi.org/10.1103/PhysRevD.81.083522}{{\em Phys.
  Rev.} {\bfseries D81} (2010) 083522},
\href{http://arxiv.org/abs/0911.3898}{{\ttfamily arXiv:0911.3898 [hep-ph]}}.

\bibitem{Tulin:2013teo}
S.~Tulin, H.-B. Yu, and K.~M. Zurek, ``{Beyond Collisionless Dark Matter:
  Particle Physics Dynamics for Dark Matter Halo Structure},''
  \href{http://dx.doi.org/10.1103/PhysRevD.87.115007}{{\em Phys. Rev.}
  {\bfseries D87} no.~11, (2013) 115007},
\href{http://arxiv.org/abs/1302.3898}{{\ttfamily arXiv:1302.3898 [hep-ph]}}.

\bibitem{Kaplinghat:2015aga}
M.~Kaplinghat, S.~Tulin, and H.-B. Yu, ``{Dark Matter Halos as Particle
  Colliders: Unified Solution to Small-Scale Structure Puzzles from Dwarfs to
  Clusters},'' \href{http://dx.doi.org/10.1103/PhysRevLett.116.041302}{{\em
  Phys. Rev. Lett.} {\bfseries 116} no.~4, (2016) 041302},
\href{http://arxiv.org/abs/1508.03339}{{\ttfamily arXiv:1508.03339
  [astro-ph.CO]}}.

\bibitem{Huo:2017vef}
R.~Huo, M.~Kaplinghat, Z.~Pan, and H.-B. Yu, ``{Signatures of Self-Interacting
  Dark Matter in the Matter Power Spectrum and the Cmb},''
  \href{http://dx.doi.org/10.1016/j.physletb.2018.06.024}{{\em Phys. Lett.}
  {\bfseries B783} (2018) 76--81},
\href{http://arxiv.org/abs/1709.09717}{{\ttfamily arXiv:1709.09717 [hep-ph]}}.

\bibitem{Bullock:2017xww}
J.~S. Bullock and M.~Boylan-Kolchin, ``{Small-Scale Challenges to the
  $\Lambda$CDM Paradigm},''
  \href{http://dx.doi.org/10.1146/annurev-astro-091916-055313}{{\em Ann. Rev.
  Astron. Astrophys.} {\bfseries 55} (2017) 343--387},
  \href{http://arxiv.org/abs/1707.04256}{{\ttfamily arXiv:1707.04256
  [astro-ph.CO]}}.

\bibitem{Fan:2010gt}
J.~Fan, M.~Reece, and L.-T. Wang, ``{Non-Relativistic Effective Theory of Dark
  Matter Direct Detection},''
  \href{http://dx.doi.org/10.1088/1475-7516/2010/11/042}{{\em JCAP} {\bfseries
  1011} (2010) 042},
\href{http://arxiv.org/abs/1008.1591}{{\ttfamily arXiv:1008.1591 [hep-ph]}}.

\bibitem{Fornengo:2011sz}
N.~Fornengo, P.~Panci, and M.~Regis, ``{Long-Range Forces in Direct Dark Matter
  Searches},'' \href{http://dx.doi.org/10.1103/PhysRevD.84.115002}{{\em Phys.
  Rev.} {\bfseries D84} (2011) 115002},
\href{http://arxiv.org/abs/1108.4661}{{\ttfamily arXiv:1108.4661 [hep-ph]}}.

\bibitem{DelNobile:2013sia}
M.~Cirelli, E.~Del~Nobile, and P.~Panci, ``{Tools for Model-Independent Bounds
  in Direct Dark Matter Searches},''
  \href{http://dx.doi.org/10.1088/1475-7516/2013/10/019}{{\em JCAP} {\bfseries
  1310} (2013) 019},
\href{http://arxiv.org/abs/1307.5955}{{\ttfamily arXiv:1307.5955 [hep-ph]}}.

\bibitem{Dent:2015zpa}
J.~B. Dent, L.~M. Krauss, J.~L. Newstead, and S.~Sabharwal, ``{General Analysis
  of Direct Dark Matter Detection: from Microphysics to Observational
  Signatures},'' \href{http://dx.doi.org/10.1103/PhysRevD.92.063515}{{\em Phys.
  Rev.} {\bfseries D92} no.~6, (2015) 063515},
\href{http://arxiv.org/abs/1505.03117}{{\ttfamily arXiv:1505.03117 [hep-ph]}}.

\bibitem{Aprile:2018dbl}
{\bfseries XENON} Collaboration, E.~Aprile {\em et~al.}, ``{Dark Matter Search
  Results from a One Ton-Year Exposure of Xenon1T},''
  \href{http://dx.doi.org/10.1103/PhysRevLett.121.111302}{{\em Phys. Rev.
  Lett.} {\bfseries 121} no.~11, (2018) 111302},
\href{http://arxiv.org/abs/1805.12562}{{\ttfamily arXiv:1805.12562
  [astro-ph.CO]}}.

\bibitem{XENON:2019gfn}
{\bfseries XENON} Collaboration, E.~Aprile {\em et~al.}, ``{Light Dark Matter
  Search with Ionization Signals in XENON1T},''
  \href{http://dx.doi.org/10.1103/PhysRevLett.123.251801}{{\em Phys. Rev.
  Lett.} {\bfseries 123} no.~25, (2019) 251801},
  \href{http://arxiv.org/abs/1907.11485}{{\ttfamily arXiv:1907.11485
  [hep-ex]}}.

\bibitem{Agnes:2018fwg}
{\bfseries DarkSide} Collaboration, P.~Agnes {\em et~al.}, ``{Darkside-50
  532-Day Dark Matter Search with Low-Radioactivity Argon},''
  \href{http://dx.doi.org/10.1103/PhysRevD.98.102006}{{\em Phys. Rev.}
  {\bfseries D98} no.~10, (2018) 102006},
\href{http://arxiv.org/abs/1802.07198}{{\ttfamily arXiv:1802.07198
  [astro-ph.CO]}}.

\end{thebibliography}\endgroup


\providecommand{\href}[2]{#2}\begingroup\raggedright\endgroup

\end{document}